\documentclass[12pt]{article}
\usepackage{amssymb}
\usepackage[pdftex]{graphicx}
\usepackage{overpic}
\usepackage{bm}
\usepackage{physics}
\usepackage[top=20truemm,bottom=20truemm,left=25truemm,right=15truemm]{geometry}
\usepackage{amsfonts,amsthm,amsmath,amssymb,upgreek}
\usepackage{units}
\usepackage{here}

\usepackage{amsfonts,amsthm,amsmath,amssymb,upgreek}
\usepackage{graphicx}
\usepackage{units}
\usepackage{float}
\usepackage [english]{babel}
\usepackage [autostyle, english = american]{csquotes}
\usepackage{framed}

\usepackage{amsfonts,amsthm,amsmath,amssymb,upgreek}
\usepackage{graphicx}
\usepackage{units}
\usepackage{float}
\usepackage [english]{babel}
\usepackage [autostyle, english = american]{csquotes}

\numberwithin{equation}{section}

\newcommand{\be}{\begin{equation}}
\newcommand{\ee}{\end{equation}}

\begin{document}
\setlength{\abovedisplayskip}{-1pt}
\setlength{\belowdisplayskip}{-1pt}

\begin{center}
{\large \bfseries Genus-Two Contributions to Late-Time Correlators in JT Gravity\par}
\vspace{1.0em}
{\normalsize Masayoshi Sato\footnote{jmasasat@nda.ac.jp} \par}
\vspace{0.5em}
{\small Department of Training, National Defense Academy of Japan\par}
\end{center}

\vspace{1.0em}

\begin{abstract}
In this work, we analyze selected genus-two ($g=2$) contributions to the late-time two-point correlation function in Jackiw--Teitelboim gravity associated with two baby universes. Extending the analysis of the single-baby-universe ($g=1$) case, we formulate an amplitude describing the emission and absorption of two baby universes during the evolution of the Hartle--Hawking state. This formulation reduces the problem to a moduli-space integral involving the Weil--Petersson volume $V_{0,3}$. To organize this integral, we decompose $V_{0,3}$ into contributions from Kontsevich graphs and extend the no-shortcut condition to a class of genus-two strip geometries. Combined with the strip approximation, this decomposition reorganizes the moduli-space integral into contributions associated with distinct geometric configurations. We explicitly evaluate contributions from the branches involving $a_1+a_2$, including a restricted contribution from a specified region of the strip parameter space. These are partial contributions to the genus-two sector of the two-point function, and the latter is evaluated only on a restricted subfamily of strip geometries.
\end{abstract}
\tableofcontents 
\newpage
\section{Introduction}
\; \; Black holes provide one of the sharpest arenas in which the tension between quantum mechanics and general relativity becomes manifest. Hawking's semiclassical calculation predicted that black holes emit thermal radiation \cite{Hawking1975}. If the semiclassical description is extrapolated throughout the evaporation process, the von Neumann entropy \cite{vonNeumann1932} of the radiation continues to increase rather than following the Page curve expected from unitary evolution \cite{Hawking1975,Hawking1976}. This behavior is in direct conflict with unitarity, a fundamental principle of quantum mechanics. If the evaporation process is unitary, the radiation entropy must eventually decrease, giving rise to the Page curve \cite{Page1993}.\\
\; \;  In recent years, our understanding of this problem has advanced significantly through the inclusion of wormhole saddle points in the gravitational path integral. New saddle points, known as replica wormholes, connect distinct replica manifolds and contribute to the entropy calculation. Their inclusion reproduces the Page curve expected from unitary evolution and thereby resolves the apparent conflict with unitarity at the level of the entropy calculation \cite{PSSY,AHMST2020}. This result demonstrates that changes in spacetime topology play a crucial role in quantum gravity. Such effects cannot be captured by conventional semiclassical calculations based on a fixed background spacetime.\\
\; \;  Jackiw--Teitelboim (JT) gravity \cite{JT1,JT2} provides a useful theoretical framework for studying topology-changing effects in quantum gravity. It is a two-dimensional dilaton gravity theory with no locally propagating bulk gravitational degrees of freedom. Accordingly, it does not support propagating gravitational waves. With fixed asymptotic AdS$_2$ boundary conditions, the remaining dynamics are captured by a boundary reparametrization mode governed by the Schwarzian action \cite{Maldacena:2016upp,Engelsoy:2016xyb,Stanford:2017thb,Mertens:2017mtv,Iliesiu:2019xuh,Malda2016,Almheiri:2014cka,Jensen:2016pah}.\\
\; \;  One of the most striking features of JT gravity is that its path integral can be evaluated with remarkable precision. The path integral involves a sum over distinct spacetime topologies. Contributions from nontrivial bulk topologies, including wormholes connecting multiple asymptotic boundaries, can be treated systematically by integrating over the corresponding moduli spaces. \\
\; \;  This structure was made precise by Saad, Shenker, and Stanford \cite{SSS}. They showed that the full genus expansion of JT gravity is reproduced by a matrix integral in the double-scaling limit. This result suggests that JT gravity is holographically dual not to a single boundary theory, but to an ensemble of quantum mechanical systems. In this matrix-model framework, wormhole contributions are naturally interpreted as statistical correlations arising from the ensemble average.\\
\; \;  Moreover, this topological expansion is closely related to the late-time behavior of the spectral form factor (SFF) \cite{SSS,Cotler:2016fpe,Saad:2018bqo}. The SFF exhibits a linearly growing ``ramp'' over an intermediate time interval. In JT gravity, this ramp is associated with a connected wormhole geometry. In the baby-universe description adopted here, the same contribution is interpreted as the emission and absorption of a single baby universe. The SFF eventually saturates to a constant value, entering the regime known as the ``plateau.'' This saturation is attributed to a nonperturbative effect reflecting the discreteness of the energy spectrum \cite{SSS,Cotler:2016fpe,Saad:2018bqo,Miyaji,TTbar}.\\
\; \;  The boundary two-point correlation function provides a powerful observable for probing the wormhole effects discussed above. In the late-time regime, it receives contributions from both the disk geometry and geometries containing a single handle. While the disk contribution decays due to the semiclassical growth of the Einstein--Rosen bridge (ERB), the one-handle contribution contains a non-decaying component. This non-decaying contribution originates from a wormhole geometry connecting the two boundaries of an eternal black hole \cite{Cotler:2016fpe,InfoMalda2001,Blommaert:2019hjr,Late,Yang}.\\
\; \;  For the late-time two-point correlation function, the genus-one ($g=1$) contribution was evaluated in Ref.\cite{M.Sato} using three independent methods. One of these methods combines the ``strip approximation'' with the ``no-shortcut condition,'' following the idea introduced in Ref.\cite{Firewall-1}. Remarkably, all three approaches give the same result and agree with the universal behavior predicted by random matrix theory. This agreement supports the validity of the strip approximation and the no-shortcut condition for describing the contribution associated with a single baby universe.\\
\; \;  In this work, we extend the analytical approach of Ref.\cite{M.Sato} to configurations involving multiple baby universes. We focus on genus-two ($g=2$) configurations involving two baby universes and analyze their contributions to the late-time two-point correlation function. A genus-two surface provides the simplest higher-genus setting beyond the genus-one case considered previously. As the genus increases, however, the structure of the moduli space becomes increasingly complicated. In particular, identifying the appropriate integration domain and avoiding the overcounting of equivalent geometries become significant technical challenges when multiple handles are present.\\
\; \;  To address these difficulties, we decompose the Weil--Petersson volume into contributions from Kontsevich graphs \cite{Koncte-1,Koncte-2,mulasepenner,norbury,ander}. We then extend the no-shortcut condition to a specific genus-two strip geometry \cite{Firewall-1,Firewall-2,Zolfi:2025},  for which we state explicitly the class of candidate spatial slices that is taken into account. Together with the strip approximation, this construction reorganizes the genus-two moduli-space integral into a finite sum of contributions associated with distinct geometric configurations. Within this framework, we explicitly evaluate the contributions associated with a specific class of genus-two strip geometries.\\
\; \;  The remainder of this paper is organized as follows. In Section 2, we review the action and partition function of JT gravity. We also discuss the Schwarzian theory governing its boundary dynamics and define the boundary two-point correlation function in the JT gravity path integral. In Section 3, following Ref.\cite{M.Sato}, we review the late-time contributions from the disk geometry and from geometries containing a single handle. In Section 4, we extend the analysis to the genus-two case. We formulate an amplitude describing the emission and absorption of two baby universes and decompose the Weil--Petersson volume $V_{0,3}$ into contributions from Kontsevich graphs. We then combine the strip approximation with the no-shortcut condition and explicitly evaluate the contributions associated with a specific class of genus-two strip geometries. Finally, in Section 5, we summarize our results and discuss their implications and possible future directions.

\vspace{-1.0em}
\section{A review of Jackiw-Teitelboim gravity}
\vspace{-1.0em}
\subsection{The Action and Partition Function of JT Gravity}
\vspace{-0.5em}
\; \;  JT gravity is a two-dimensional model of quantum gravity that admits explicit black hole solutions. It has been extensively studied as a tractable framework for analyzing gravitational path integrals and quantum effects in black hole physics. Its action is given by \cite{JT1,JT2,Muta1992,Nojiri:2024ycf}\\
\begin{align}
 I_{\text{JT}} = -\frac{1}{16\pi G_N^{(2)}} \int_{\Sigma}\sqrt{g}\phi(R+2)-\frac{1}{8\pi G_N^{(2)}}\int_{\partial\Sigma}\sqrt{h}\phi (K-1), \label{2-1} \\ \nonumber
\end{align}
where $G_N^{(2)}$ is the two-dimensional Newton constant, $\Sigma$ is a two-dimensional Riemann surface equipped with a metric $g$, and $\phi$ is the dilaton field. Here, $R$ denotes the scalar curvature in the interior of $\Sigma$, $K$ is the extrinsic curvature of the boundary $\partial\Sigma$, and $h$ is the induced metric on the boundary.\\
\; \;  A central feature of JT gravity is that the dilaton $\phi$ acts as a Lagrange multiplier. Varying the action with respect to $\phi$ imposes $R=-2$. The bulk spacetime is therefore locally AdS$_2$ \cite{Maldacena:2016upp} and has no locally propagating gravitational degrees of freedom. Under standard asymptotic AdS$_2$ boundary conditions, the remaining dynamics are encoded in a boundary time-reparametrization mode governed by the Schwarzian action:\\
\begin{align} 
I_{\text{Sch}}[F] = - \int^{\beta}_{0} d\tau \text{Sch}(F(\tau),\tau), \label{2-2}
\end{align}
\\
where $F(\tau)$ is the boundary time-reparametrization mode, $\tau$ is the Euclidean time, and $\beta$ is the inverse temperature of the black hole. $ \mathrm{Sch}(F(\tau), \tau)$ denotes the Schwarzian derivative.\\
\; \;  For the disk geometry, which represents the simplest saddle-point topology, the gravitational path integral reduces to the path integral of the Schwarzian theory \cite{SSS}. For spacetimes with nontrivial topology, such as wormhole geometries, the path integral also includes an integration over the moduli characterizing the bulk geometry. Accordingly, the partition function for a Riemann surface of genus $g$ with $n$ asymptotic boundaries can be written schematically as \\
\begin{align}
Z_{g,n}(\beta_1,\dots,\beta_n) = e^{S_{0}(2-2g-n)}\int_{\mathcal{M}_{g,n}} d(\text{moduli}) \ \prod^{n}_{i=1} \int_{\text{Diff}(S^1)}\mathcal{D}F_i e^{- \int^{\beta_i}_0 d\tau_i \text{Sch}(F_i,\tau_i)}, \label{2-3} \\ \nonumber
\end{align}
where $\mathcal{M}_{g,n}$ is the moduli space of genus-$g$ Riemann surfaces with $n$ boundaries, $\beta_i$ is the inverse temperature assigned to the $i$-th asymptotic boundary, and $S_0$ denotes the extremal entropy of the black hole. The symbol $d (\mathrm{moduli})$ represents the measure on the corresponding moduli space. The JT gravity path integral thus provides a natural framework for treating black hole thermodynamics and topology-changing quantum gravitational effects, including wormhole contributions, within a unified description \cite{SSS,Review-1}.
\vspace{-0.5em}
\subsection{Two-point correlation function in JT gravity}
\begin{figure}[H]
\centering
\includegraphics[width=160mm]{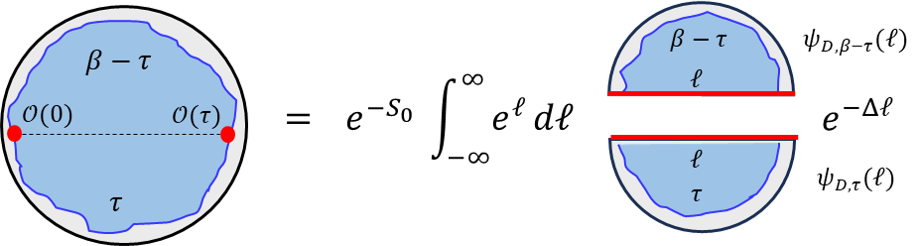}
\caption{\small Disk contribution to the two-point correlation function in Euclidean signature.}
\label{fig3-1(corr-11)}
\end{figure}

\begin{figure}[H]
\centering
\includegraphics[width=150mm]{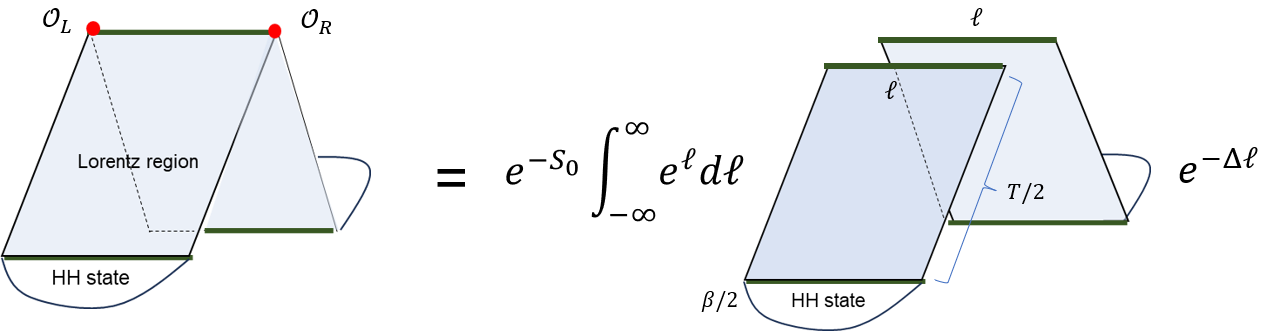}
\caption{\small Disk contribution to the two-point correlation function in Lorentzian signature.}
\label{fig3-2(corr-2)}
\end{figure}

\begin{figure}[H]
\centering
\includegraphics[width=160mm]{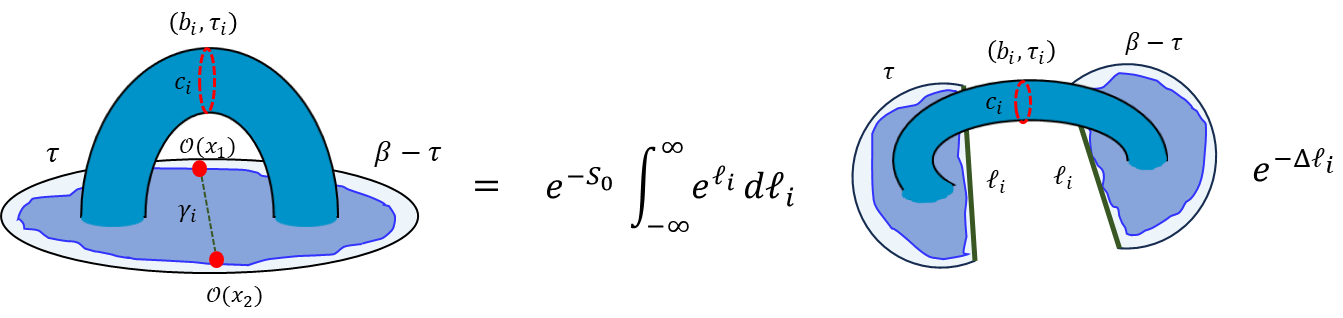}
\caption{\small Cutting the geometry along a geodesic $\gamma_i$ passing through the handle yields a double-trumpet hyperbolic surface. The corresponding contribution can be expressed in terms of two trumpet wavefunctions $\psi_{\mathrm{Tr}}(\ell,b)$.}
\label{fig3-ramp-1}
\end{figure}

\vspace{-0.5em}
\; \;  Boundary two-point correlation functions provide basic probes of quantum gravitational dynamics in JT gravity. In the AdS/CFT framework \cite{AdS/CFT-1,AdS/CFT-2}, boundary operators encode information about the bulk geometry and the corresponding quantum state. Their correlation functions can therefore be used to investigate both bulk quantum effects and real-time evolution.\\
\; \;  For boundary operators $\mathcal{O}(x_i)$ of conformal dimension $\Delta$, inserted at boundary points $x_i$, the two-point correlation function is defined by combining the JT gravity path integral with the matter correlation function \cite{Late,Yang}:\\
\begin{align}
\langle \mathcal{O}(x_1)\mathcal{O}(x_2)\rangle_{\text{JT}} = \int \mathcal{D}g\ \mathcal{D}\phi \, e^{-I_{\text{JT}}[g,\phi]} \, \langle \mathcal{O}(x_1)\mathcal{O}(x_2)\rangle_{\text{CFT}}. \label{2-4} 
\end{align}
\\
\; \;  We assume that the boundary operators do not couple directly to the dilaton field. In the following, we examine the disk contribution and the contribution from a disk with one handle in a two-sided black hole geometry.\footnote{A two-sided black hole is described by an eternal AdS$_2$ geometry with two asymptotic boundaries. The left and right boundaries correspond to two independent quantum systems, or equivalently two copies of the boundary theory, whose joint state is the entangled thermofield double (TFD) state. On the gravity side, this entanglement is geometrically represented by an ERB connecting the two asymptotic regions \cite{InfoMalda2001,Israel:1976ur,VanRaamsdonk:2010pw}.}
\paragraph{$\blacksquare$ Disk ($g=0$) contribution}
The simplest topology is the disk, whose Euler characteristic is $\chi=1$. Let $\ell$ denote the length of the geodesic connecting the two operator insertions, $\mathcal{O}(x_1)$ and $\mathcal{O}(x_2)$. The propagation of the matter field between these insertion points contributes a factor $e^{-\Delta\ell}$ \cite{Late,Yang}. As shown in Ref.~\cite{Yang}, (\ref{2-4}) can be evaluated by cutting the spacetime along this geodesic. A Hartle--Hawking (HH) wavefunction \cite{HH-1,HH-2} is associated with each of the two resulting Riemann surfaces. The original geometry is then reconstructed by gluing the two surfaces along the geodesic and inserting the factor $e^{-\Delta\ell}$, as illustrated in Fig.~\ref{fig3-1(corr-11)}. The disk contribution to the two-point correlation function is therefore given by \\
\begin{align}
\langle\mathcal{O}(\tau)\mathcal{O}(0)\rangle_{\chi=1} &= e^{-S_0} \int^{\infty}_{-\infty} e^\ell \ d\ell \ \psi_{D,\beta-\tau} (\ell) \psi_{D, \tau}(\ell) e^{-\Delta \ell}  \label{2-5} \\  
&= e^{-S_0} \int^{\infty}_{0} dE_1dE_2 \ \rho_0(E_1)\rho_0(E_2)\ e^{-(\beta-\tau)E_1-\tau E_2} |\mathcal{O}_{E_1,E_2}|^{2}, \label{2-6} \\ \nonumber 
\end{align}
\noindent where we have set $x_1=\tau$ and $x_2=0$. The leading-order density of states in JT gravity is $\rho_0(E)=\frac{e^{S_0}}{2\pi^2}\sinh\left(2\pi\sqrt{2E}\right)$, and $\mathcal{O}_{E_1,E_2}$ denotes the matrix element of the boundary operator between energy eigenstates with energies $E_1$ and $E_2$ \cite{Review-1}.\\
\; \;  The corresponding real-time correlator is obtained by analytically continuing the Euclidean time variable $\tau$ in (\ref{2-6}) to Lorentzian signature \cite{Late}. For the two-sided two-point correlation function, the appropriate continuation is $\tau \to \frac{\beta}{2}+iT$. Under this continuation, the two disk wavefunctions become complex conjugates of each other. The resulting disk contribution, illustrated in Fig.\ref{fig3-2(corr-2)}, is given by \\
\begin{align}
\langle\mathcal{O}(\beta/2+iT)\mathcal{O}(0)\rangle_{\chi=1}&= e^{-S_0} \int^{\infty}_{-\infty} e^{\ell} d\ell \ |\psi_{D,\beta/2+ i T}(\ell)|^2 e^{-\Delta \ell} \nonumber \\
&= e^{-S_0} \int^{\infty}_{0} dE_1dE_2 \ \rho_0(E_1)\rho_0(E_2)\ e^{-\frac{\beta}{2}(E_1+E_2)-iT(E_1-E_2)} |\mathcal{O}_{E_1,E_2}|^{2}. \label{2-7}
\end{align}
\vspace{-1.0em}
\vspace{-2.2em}
\paragraph{$\blacksquare$ \ Disk with one handle ($g=1$) contribution}
For the disk with one handle, whose Euler characteristic is $\chi=-1$, there are infinitely many homotopically inequivalent geodesics connecting the two operators. The corresponding two-point correlation function is therefore considerably more involved than in the disk case. As discussed in Refs.\cite{Late,Volume}, on a fixed hyperbolic geometry with one handle, the matter two-point function is given by a sum over the geodesics representing the homotopy classes connecting the two operator insertions:\\
\begin{align}
\langle \mathcal{O}(x_1)\mathcal{O}(x_2) \rangle_{\text{CFT}} = \sum_{\gamma \in \mathcal{G}_{x_1,x_2}} e^{-\Delta \ell_\gamma}. \label{2-8} 
\end{align}
\\
where $\mathcal{G}_{x_{1},x_{2}}$ denotes the set of geodesic representatives of all homotopy classes connecting the boundary points $x_1$ and $x_2$.\\
\; \; Two different constructions are used in this paper, and we distinguish them explicitly. (\ref{2-8}) is the sum over geodesics in all homotopy classes connecting the two operators, with weight $e^{-\Delta\ell_\gamma}$; it gives the matter correlator on a fixed geometry. (\ref{2-9}) is the unfolded form of this sum \cite{Late}: the sum over homotopy classes converts the moduli-space integral into an integral over a fundamental domain. By contrast, in Secs.3.2 and 4 we follow the prescription of Stanford and Yang \cite{Firewall-1}. For each geometry we select a single physical spatial slice, namely the shortest admissible geodesic connecting the two boundary points, and restrict the moduli integral to the region where this slice is the $\ell$-geodesic (the no-shortcut condition). Ref.\cite{M.Sato} showed that the three approaches considered there agree at genus one and reproduce the late-time ramp. In the $\Delta\to0$ limit the geodesic sum diverges, whereas the restricted integral remains finite. We use ``contribution" in Secs.3.2 and 4 in the latter sense. \\
\;\; We now evaluate the contribution of the disk with one handle to the late-time observable, denoted by $\langle \mathcal{O}(x_1)\mathcal{O}(x_2) \rangle_{\chi=-1}^{\rm handle}$. The nontrivial contribution arises from geometries in which the selected shortest admissible geodesic connecting the two boundary points passes through the handle without self-intersection. \\
\; \;  As shown in Fig.\ref{fig3-ramp-1}, cutting the disk with one handle along the geodesic $\gamma_i$ produces a connected hyperbolic surface whose genus is reduced by one. This surface can be interpreted as a double trumpet with two asymptotic boundaries \cite{Late}. It contains a closed geodesic $c_i$ that is disjoint from $\gamma_i$. A further cut along $c_i$ decomposes the double trumpet into two trumpet geometries.
\; \;  Each homotopically inequivalent geodesic $\gamma_i$ labels a distinct mapping-class-group image of a fundamental domain in Teichm\"uller space. Summing over all such geodesics unfolds the moduli-space integral into an integral over the corresponding covering space. After imposing the Dehn-twist identification, the contribution can be represented by an integral over the fundamental domain $0<b<\infty$ and $0<\tau<b$:\\
\begin{align}
\langle \mathcal{O}(x_1)\mathcal{O}(x_2) \rangle_{\chi=-1}^{\text{handle}}  = e^{-S_0} \int_0^\infty db \int^{b}_{0} d\tau \int_{-\infty}^\infty e^\ell d\ell \ \psi_{\text{Tr},\tau}(\ell,b) \psi_{\text{Tr},\beta-\tau}(\ell,b) e^{-\Delta \ell}. \label{2-9}
\end{align}
\vspace{-2.0em}
\vspace{-1.5em}
\noindent \; \;  Unlike the disk contribution discussed above, the contribution from the disk with one handle contains a component that does not decay at late times. This non-decaying component is associated with the ramp behavior of the SFF.
\vspace{-1.0em}
\subsection{Properties of JT gravity at late times}
\begin{figure}[H]
\centering
\includegraphics[width=130mm]{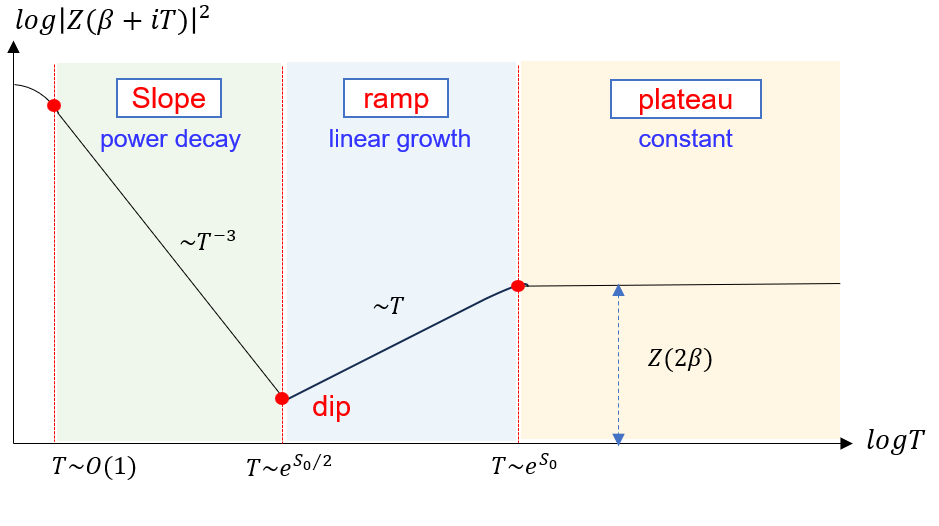}
\caption{\small Schematic behavior of the SFF in JT gravity.}
\label{fig2-1(SFF)}
\end{figure}

\; \;  In JT gravity, effects beyond the classical description become particularly important in the late-time regime $T\sim O(e^{S_0})$. These effects are reflected in the spectral form factor (SFF), which probes correlations in the energy spectrum through its real-time evolution. The SFF initially exhibits a power-law decay known as the ``slope,'' followed by a regime of linear growth called the ``ramp.'' At the Heisenberg time, $T_H=2\pi\rho_0(E)$, the ramp terminates, and the SFF approaches a constant value known as the ``plateau'' (see Fig.~\ref{fig2-1(SFF)}) \cite{SSS,Cotler:2016fpe,Saad:2018bqo,Stanford:2019vob}.
This behavior can be described using the universal spectral density correlation function of the random matrix model dual to JT gravity \cite{SSS,Stanford:2019vob}:\\
\begin{align}
\left\langle \rho(E_1)\rho(E_2)\right\rangle & \approx
\rho_0(E_1)\rho_0(E_2) -\frac{\sin^2(\pi \rho_0(E_1)(E_1-E_2))}{\pi^2(E_1-E_2)^2} + \rho_0(E_1) \delta(E_1-E_2).  \label{2-10} \\ \nonumber
\end{align} 
The second term contains the sine kernel and describes the universal repulsion between nearby energy levels in quantum-chaotic systems.\\
\; \;  From the perspective of JT gravity, the three terms in (\ref{2-10}) have distinct interpretations. The disconnected product in the first term corresponds to two independent disk geometries and gives rise to the slope contribution to the SFF. The sine-kernel term is associated with the connected double-trumpet geometry, which represents a wormhole connecting the two boundaries. This contribution gives rise to the ramp behavior of the SFF. The delta-function term reflects the discreteness of the energy spectrum and gives rise to the plateau. Unlike the disk and double-trumpet contributions, the plateau is a nonperturbative effect that cannot be captured by the perturbative genus expansion.
\vspace{-1.0em}
\section{\large Evaluation of the Two-Point Correlation Function at Late Times}
\vspace{-0.5em}
\; \;  This section reviews the analysis of the late-time two-point correlation function presented in Ref.\cite{M.Sato}. In that work, the contributions from the disk geometry and the disk with one handle were evaluated using three independent approaches.\\
\; \;  The second approach builds on the method introduced in Ref.\cite{Firewall-1}. It evaluates the correlation function by integrating over a region of moduli space selected by the no-shortcut condition. Although this approach is approximate, the strip approximation based on ribbon graphs makes it possible to identify the relevant integration domain within the otherwise complicated moduli space. The use of the strip approximation is motivated by the fact that the relevant geometries become highly elongated in the late-time regime. (See the remark after (\ref{2-8}) for the distinction between the geodesic sum and the Stanford–Yang slice-selection prescription used here.)\\
\; \;  The three approaches were shown in Ref.\cite{M.Sato} to yield consistent results. These results also agree with the SFF analysis reviewed in the previous section. This agreement supports the validity of the strip approximation combined with the no-shortcut condition and suggests that the same framework can be applied to more complicated higher-genus geometries. The discussion in this section provides the basis for the genus-two analysis presented in the next section.
\vspace{-1.0em}
\subsection{Disk ($g=0$) contribution}
\; \;  After the analytic continuation $\tau \to \beta/2+iT$ in (\ref{2-6}), the disk contribution to the two-point correlation function takes the form:\\
\begin{align}
\langle \mathcal{O}(\beta/2+iT) \mathcal{O}(0)\rangle_{\chi=1}  & = e^{-S_0} \int^{\infty}_{-\infty}e^{\ell}d\ell e^{-\Delta \ell} \ \psi_{D,\beta/2+iT}(\ell)\psi_{D,\beta/2-iT}(\ell) \label{3-1} \\
&= e^{-S_0} \int^{\infty}_{-\infty}e^{\ell}d\ell e^{-\Delta \ell} \int^{\infty}_{0} dE_1dE_2 \ e^{-\frac{\beta}{2}(E_1+E_2)-iT(E_1-E_2) } \rho_0(E_1)\rho_0(E_2) \psi_{E_1}(\ell)\psi_{E_2}(\ell). \label{3-2}
\end{align} 
\\
Applying an inverse Laplace transform with respect to $\beta$, we obtain the corresponding microcanonical correlator at fixed energy $E$:\\
\begin{align}
G(E,T)_{\chi=1}&\equiv \int_{\mathcal{C}} \frac{d\beta}{2\pi i} e^{\beta E} \langle \mathcal{O}(\beta/2+iT) \ \mathcal{O}(0)\rangle_{\chi=1} \nonumber \\
&= e^{-S_0}\int^{\infty}_{-\infty}e^{\ell}d\ell e^{-\Delta \ell}  \int_{\mathcal{C}} \frac{d\beta}{2\pi i} e^{\beta E} \psi_{D,\beta/2+iT}(\ell)\psi_{D,\beta/2-iT}(\ell). \label{3-3} \\ \nonumber  
\end{align}
The integration over $\beta$ imposes the constraint $E_1+E_2=2E$. We therefore parameterize the two energies as
\begin{align}
E_1=E+\frac{\omega}{2}, \hspace{20pt} E_2=E-\frac{\omega}{2}. \label{3-4} 
\end{align}
\\
Using this parametrization, the expression in (\ref{3-3}) can be rewritten as an integral over $\omega$:
\\
\begin{align}
G(E,T)_{\chi=1} = e^{-S_0}\int^{\infty}_{-\infty}e^{\ell}d\ell e^{-\Delta \ell}  \int^{\infty}_{-\infty} d\omega e^{-iT\omega } \rho_0(E_1)\rho_0(E_2)\psi_{E_1}(\ell)\psi_{E_2}(\ell). \label{3-5} 
\end{align}
\\
In evaluating the integral, we employ the semiclassical approximation used in Ref.\cite{Firewall-1} and assume that $E,T\gg 1$. In this regime, the modified Bessel function $K_{2i\sqrt{2E}}(4e^{-\ell/2})$ appearing in $\psi_E(\ell)$ is exponentially suppressed for $\ell<-\log E$ and rapidly oscillates for $\ell>-\log E$.\\
\; \;  In the oscillatory region, the Bessel function can be approximated as \cite{bessal}\\
\begin{align}
&e^{\ell/2}\psi_{E}(\ell) = 4K_{2i\sqrt{2E}}(4e^{-\ell/2}) \approx \frac{4\pi^{1/2}}{(2E)^{1/4}} e^{-\pi\sqrt{2E}} \cos \left[\ \sqrt{2E} (\ell+\log(2E)-2) -\frac{\pi}{4} \ \right]. \label{3-6} \\ \nonumber
\end{align}
Substituting (\ref{3-6}) into (\ref{3-5}) and performing the integral over $\omega$, we obtain two delta functions:\\
\begin{align}
G(E,T)_{\chi=1} \approx \frac{e^{S_0+2\pi\sqrt{2E}}}{2\pi^2(2E)^{1/2}}\int^{\infty}_{-\log E} d\ell e^{-\Delta \ell}  \ \left[ \delta \left(T-\frac{L}{\sqrt{2E}} \right) + \delta \left(T+\frac{L}{\sqrt{2E}} \right) \right], \label{3-7} \\ \nonumber 
\end{align}
where $L$ is defined by $L=\ell+\log(2E)-2$.\\
\; \;  For $T\gg 1$, the second delta function has no support in the relevant integration region and can therefore be discarded. The disk contribution is then given by\\
\begin{align}
G(E,T)^{\chi=1}_{\text{no-baby}} \approx \frac{e^{S_0+2\pi\sqrt{2E}}}{2\pi^2} \int^{\infty}_{-\log E} d\ell e^{-\Delta \ell} \delta(\ell-\ell_T) = \frac{e^{S_0+2\pi\sqrt{2E}}}{2\pi^2} e^{-\Delta \ell_T}, \label{3-8} \\ \nonumber
\end{align}
where $\ell_T$ is defined by $\ell_T=\sqrt{2E}T-\log(2E)+2 \sim \sqrt{2E}T$. Thus, the disk contribution to the two-point correlation function is localized at the geodesic length $\ell=\ell_T$. In the late-time regime, $\ell_T$ becomes large, and the factor $e^{-\Delta \ell_T}$ exponentially suppresses the disk contribution.\\
\; \;  Since the disk geometry has no moduli, the three approaches considered in Ref.\cite{M.Sato} all yield the same result, given in (\ref{3-8}).
\vspace{-1.0em}
\subsection{Disk with one handle ($g=1$) contribution}
\begin{figure}[H]
\centering
\includegraphics[width=185mm]{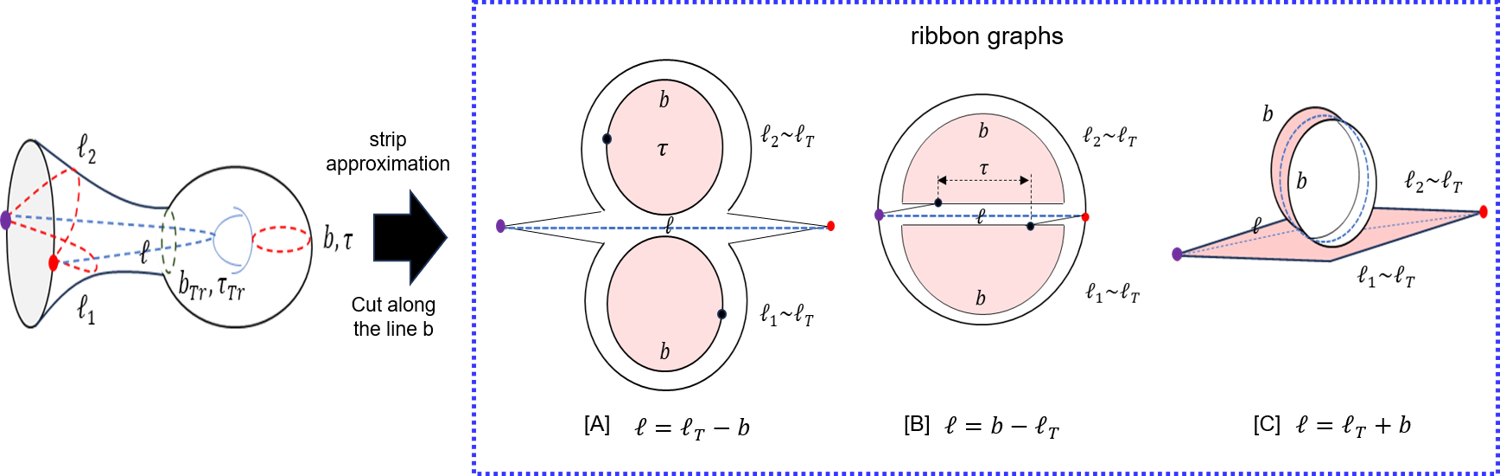}
\caption{\small Depending on the value of the twist parameter $\tau$, there may exist a geodesic connecting the boundary operators that is shorter than the $\ell$-geodesic indicated by the dashed line in the figure. In such a case, the $\ell$-geodesic cannot be regarded as an admissible spatial slice. Such configurations are excluded by imposing the no-shortcut condition, which restricts the moduli parameters $b$ and $\tau$ so that the $\ell$-geodesic defines the shortest admissible spatial slice.}
\label{fig5-2(strip)} 
\end{figure}

\begin{figure}[H]
\centering
\includegraphics[width=100mm]{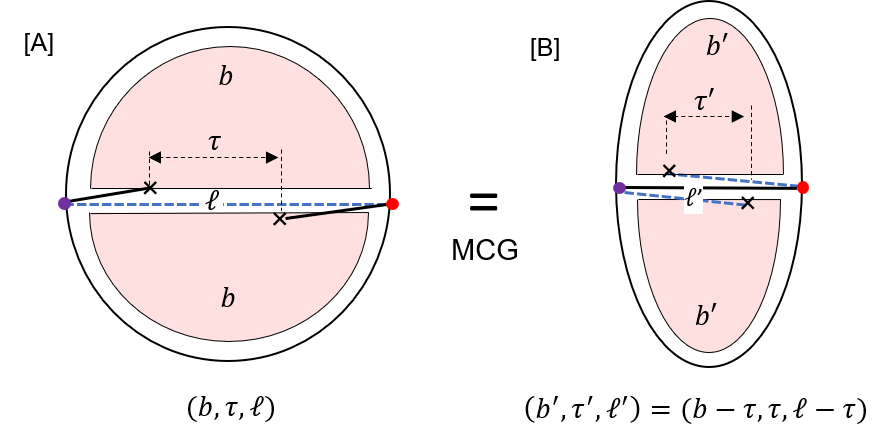}
\caption{\small [A] Configuration in which the no-shortcut condition is violated. [B] A mapping class group transformation maps the $\ell$-geodesic to an $\ell'$-geodesic whose length is shorter than that of the original $\ell$-geodesic by $\tau$. The admissible spatial slice is then given by the $\ell'$-geodesic rather than the original $\ell$-geodesic.}
\label{fig5-3(strip-2)}
\end{figure}

\vspace{-0.5em}
\; \;  As in the disk case discussed in the previous subsection, we evaluate the one-handle contribution in the regime $E,T\gg 1$. After the analytic continuation $\tau \to \beta/2+iT$ in (\ref{2-9}), the contribution from the disk with one handle takes the form\\
\begin{align}
& \langle \mathcal{O}(\beta/2+iT) \mathcal{O}(0)\rangle_{\chi=-1} \nonumber \\
& = e^{-S_0} \int^{\infty}_{-\infty}e^{\ell}d\ell e^{-\Delta \ell}  \int d\tau db \int^{\infty}_{0} dE_1dE_2 \ e^{-\frac{\beta}{2}(E_1+E_2)-iT(E_1-E_2) } \rho_{\text{Tr}}(E_1,b)\rho_{\text{Tr}}(E_2,b) \psi_{E_1}(\ell)\psi_{E_2}(\ell). \label{3-9} 
\end{align} 
\\
As in the previous subsection, we define $G(E,T)_{\chi=-1}$ by an inverse Laplace transform. We then parametrize the energies as in (\ref{3-4}) and evaluate the wavefunctions $\psi_{E_i}(\ell)$ using (\ref{3-6}). The subsequent integration over $\omega$ yields four delta functions:\\
\begin{align}
G(E,T)_{\chi=-1} & = \frac{2e^{-S_0-2\pi \sqrt{2E}}}{(2E)^{3/2}} \ \int^{\infty}_{-\log E} d\ell e^{-\Delta \ell}\int d\tau db \nonumber \\
&\times \left[ \delta \left(T-\frac{(L+b)}{\sqrt{2E}} \right) + \delta \left(T-\frac{(L-b)}{\sqrt{2E}} \right) + \delta \left(T+\frac{(L-b)}{\sqrt{2E}} \right)  + \delta \left(T+ \frac{(L+b)}{\sqrt{2E}} \right)  \right] \nonumber \\
& = \frac{e^{-S_0-2\pi \sqrt{2E}}}{E} \int^{\infty}_{-\log E} d\ell e^{-\Delta \ell}\int db d\tau \ \delta(\ell \pm \ell_{T} \pm b). \label{3-10} 
\end{align}
\\
For $T\gg 1$ and $b>0$, $\delta(\ell+\ell_T+b)$ has no support in the allowed region and therefore does not contribute. The remaining three delta functions correspond to distinct geometric configurations in the strip approximation\footnote{The strip approximation is motivated by the Gauss--Bonnet theorem. For hyperbolic surfaces with geodesic boundaries, the area is fixed by the Euler characteristic. In the late-time regime, the relevant geodesic boundaries become large, and the surface is well approximated by elongated strip-like regions. In this limit, the geometry can be represented by a ribbon graph. This description greatly simplifies moduli-space integrals, especially for higher-genus surfaces \cite{Firewall-1,Firewall-2,Koncte-1}.}.\\
\; \;  In Ref.\cite{M.Sato}, the geometries associated with the remaining delta functions are represented by ribbon graphs. The corresponding contributions to the two-point correlation function are then evaluated explicitly by imposing the ``no-shortcut condition''\footnote{In Ref.\cite{Firewall-1}, for the case $g=1$, the partition function is evaluated by integrating over the region of moduli space selected by the no-shortcut condition. This condition excludes configurations in which two points on the chosen geodesic are connected by a shorter path through the spacetime. It also prevents overcounting of the same surface by restricting the integration domain in moduli space.\\
\; \;  This construction follows from the fact that the Teichm\"uller space 
$\mathcal{T}_{\Sigma_{1,1}}$ is a covering space of the moduli space $\mathcal{M}_{\Sigma_{1,1}}$. Homotopically distinct geodesics connecting the two operators correspond to different fundamental domains in $\mathcal{T}_{\Sigma_{1,1}}$. For example, different geodesics $\gamma_1$ and $\gamma_2$ passing through the handle belong to the same cutting class. The regions in which each geodesic defines the shortest spatial slice are related by the action of the mapping class group. The no-shortcut condition can therefore be interpreted as restricting the integration to one of these regions. There also exist regions in which geodesics that do not pass through the handle define the shortest spatial slice.}. This condition ensures that the chosen geodesic defines the shortest admissible spatial slice. Among the three remaining configurations, we focus on those associated with $\delta(\ell-\ell_T+b)$ and $\delta(\ell+\ell_T-b)$.
\vspace{-1.0em}
\paragraph{$\blacksquare$ \ $\delta(\ell-\ell_T+b)$ } 
This configuration is shown in Fig.\ref{fig5-2(strip)}[A]. In this configuration, the no-shortcut condition imposes no additional constraint on either the baby-universe size $b$ or the twist parameter $\tau$. This is because the geodesic of length $\ell$, which we call the $\ell$-geodesic, is always shorter than the other candidates, namely the $\ell_1$- and $\ell_2$-geodesics. The $\ell$-geodesic is therefore selected as the spatial slice.\\
\; \;  Thus, the twist parameter is integrated only over the fundamental domain associated with the Dehn twist identification, $0<\tau<b$. The condition $\ell=\ell_T-b>-\log E$, together with $b>0$, gives $0<b<\ell_T+\log E$ and $-\log E<\ell<\ell_T$.
As illustrated in Fig.\ref{fig5-2(strip)}[A], geodesics longer than $\ell_T$ are not selected as spatial slices. This explains the upper bound $\ell<\ell_T$:\\
\begin{align}
G(E,T)^{\ell=\ell_T-b}_{\chi=-1} &= \frac{e^{-S_0-2\pi \sqrt{2E}}}{E} \int^{\ell_T}_{-\log E} d\ell e^{-\Delta \ell}\int^{\ell_T+\log E}_{0} db \int^{b}_{0}d\tau \delta(\ell-\ell_T+b) \nonumber \\
& = \frac{e^{-S_0-2\pi \sqrt{2E}}}{E} \left[\ \left( \ell_T + \log E - \frac{1}{\Delta} \right)\frac{E^{\Delta}}{\Delta} + \frac{e^{-\Delta \ell_T}}{\Delta^{2}}  \ \right]. \label{3-11} \\ \nonumber
\end{align}
\noindent \; \;  To examine the late-time behavior of (\ref{3-11}), we take the $\tau$-scaling limit, $T,\,S_0\to\infty$, with $Te^{-S_0}$ held fixed \cite{tau,OS}. This limit isolates the contribution associated with the ramp behavior of the SFF:\\
\begin{align}
&\frac{e^{-S_0-2\pi \sqrt{2E}}}{E} \left[\ \left( \ell_T + \log E- \frac{1}{\Delta} \right)\frac{E^{\Delta}}{\Delta} + \frac{e^{-\Delta \ell_T}}{\Delta^{2}}  \ \right] \ \underset{T \to \infty} \rightarrow \ \frac{\sqrt{2}e^{-S_0-2\pi \sqrt{2E}}}{\Delta} TE^{\Delta-\frac{1}{2}}. \label{3-12} \\ \nonumber
\end{align}
In Ref.\cite{M.Sato}, all three approaches were shown to give consistent results and to reproduce the ramp behavior of the SFF in this regime.
\vspace{-1.0em}
\paragraph{$\blacksquare$ \ $\delta(\ell+\ell_T-b)$}
This configuration is shown in Fig.\ref{fig5-2(strip)}[B]. In this case, the no-shortcut condition must be imposed to ensure that the selected geodesic defines the shortest admissible spatial slice. For $0<\tau<\ell$, the twist identification relates two points on the ribbon graph separated by $\tau$, as shown in Fig.\ref{fig5-3(strip-2)}. These two points can then be connected through the spacetime by a path shorter than the corresponding segment of the $\ell$-geodesic. Consequently, the $\ell$-geodesic no longer defines the shortest spatial slice and cannot be regarded as admissible in this region.\\
\; \;  To identify an admissible representative of the same geometry, we apply a mapping class group transformation. As shown in Fig.\ref{fig5-3(strip-2)}[B], this transformation maps the $\ell$-geodesic to an $\ell'$-geodesic whose length is shorter by $\tau$. In the strip approximation, the transformation acts on the relevant parameters as $(b,\tau,\ell)\longrightarrow (b-\tau,\tau,\ell-\tau)$. When $\ell>|\tau|$, the transformation can be applied repeatedly to obtain a shorter geodesic. Once $\ell<|\tau|$, no further shortening is possible. The integration domain is chosen to be the region in which this condition is satisfied. In this region, the resulting $\ell'$-geodesic defines the shortest admissible spatial slice \cite{Firewall-1}.\\
\; \;  After imposing the no-shortcut condition, the relation $\ell=b-\ell_T$ gives the allowed range $b-\ell_T<\tau<\ell_T$ for the twist parameter. In addition, the conditions $\ell=b-\ell_T>-\log E$ and $\ell<\ell_T$ imply $\ell_T-\log E<b<2\ell_T$. Using these integration ranges, the contribution is evaluated as\\
\begin{align}
G(E,T)^{\ell=b-\ell_T}_{\chi=-1}  &= \frac{e^{-S_0-2\pi \sqrt{2E}}}{E} \int^{\ell_T}_{-\log E} d\ell e^{-\Delta \ell}\int^{2\ell_T}_{\ell_T-\log E} db \int^{\ell_T}_{b-\ell_T}d\tau \delta(\ell+\ell_T-b) \nonumber \\
& = \frac{e^{-S_0-2\pi \sqrt{2E}}}{E} \left[\ \left( \ell_T + \log E- \frac{1}{\Delta} \right)\frac{E^{\Delta}}{\Delta} + \frac{e^{-\Delta \ell_T}}{\Delta^{2}}  \ \right]. \label{3-13}  \\ \nonumber
\end{align}
Although the no-shortcut condition restricts the integration domain in moduli space, the final result coincides with that obtained for the configuration shown in Fig.\ref{fig5-2(strip)}[A]. Taking the $\tau$-scaling limit again yields the same late-time behavior as in (\ref{3-12}).
\vspace{-1.0em}
\paragraph{$\blacksquare$ \ $\delta(\ell-\ell_T-b)$ } This configuration is shown in Fig.\ref{fig5-2(strip)}[C]. As discussed in Ref.\cite{Firewall-1}, no region of the moduli space associated with this ribbon graph satisfies the no-shortcut condition. For any choice of $b$ and $\tau$, the $\ell$-geodesic is longer than both the $\ell_1$- and $\ell_2$-geodesics because it detours around the $b$-cycle. The spatial slice connecting the two operators is therefore defined by the shorter of the $\ell_1$- and $\ell_2$-geodesics. Consequently, this ribbon graph does not give an independent leading one-handle contribution. Instead, it contributes only a correction to the disk result in (\ref{3-8}) and does not affect the leading late-time behavior of the correlation function. The corresponding subleading contribution is given by\\
\begin{align}
G(E,T)^{\ell=b+\ell_T}_{\chi=-1}  &= \frac{e^{-S_0-2\pi \sqrt{2E}}}{E} \int^{\infty}_{-\log E} d\ell e^{-\Delta \ell}\int^{\infty}_{0} db \int^{b}_{0}d\tau \delta(\ell-\ell_T-b) \nonumber \\
& = \frac{e^{-S_0-2\pi \sqrt{2E}}}{\Delta^2 E} e^{-\Delta \ell_T} \ \underset{T,S_0 \to \infty} \rightarrow 0. \label{3-14} 
\end{align}
\vspace{-1.0em}
\section{Extension to Genus-Two Surfaces} \label{g=2 correlator}
\vspace{-0.8em}
\; \;  So far, we have focused on spacetime geometries with a single handle and analyzed their contributions to the two-point correlation function \cite{M.Sato}. These contributions were interpreted as effects associated with a single baby universe. In this section, we extend the analysis to geometries with two handles, corresponding to configurations involving two baby universes. A direct integration over the genus-two moduli space is technically challenging because one must identify the appropriate integration domain while avoiding the overcounting of equivalent geometries. In the late-time regime, this difficulty can be addressed by reorganizing the moduli-space integral in terms of strip geometries. The no-shortcut condition then restricts the moduli parameters so that the chosen geodesic defines the shortest admissible spatial slice. This construction provides a tractable framework for explicitly evaluating a class of genus-two contributions.
\vspace{-1.0em}
\subsection{\normalsize{The Amplitude for the Emission and Absorption of Two Baby Universes}}
\begin{figure}[H]
\centering
\includegraphics[width=90mm]{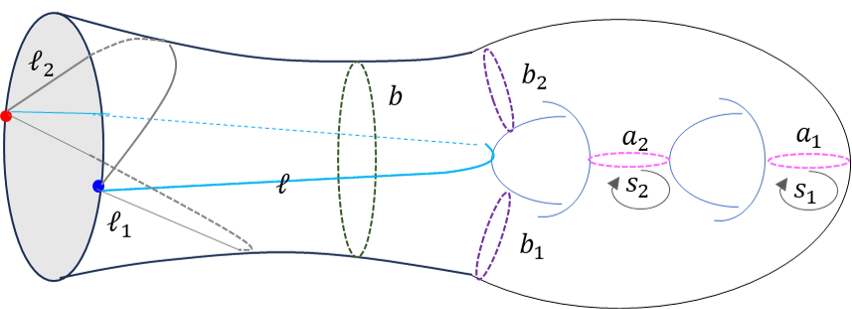}
\caption{\small Parameter configuration for the two-point correlation function in (\ref{4-1}). The $\ell$-geodesic is chosen to pass through both handles.}
\label{fig7-1-g=2}
\end{figure}

\vspace{-0.5em}
\; \;  Motivated by the preceding discussion, we formulate the contribution involving two baby universes as an amplitude describing their emission and absorption during the evolution of the Hartle--Hawking state:\\
\begin{align}
\langle \mathcal{O}(x_1)\mathcal{O}(x_2) \rangle_{\chi=-3} &= e^{-3S_{0}} \int_{-\infty}^{\infty} e^{\ell}e^{-\Delta \ell} d\ell \int_{-\infty}^{\infty} \prod_{i=1}^{2} e^{\ell_i}d \ell_{i} \int \prod^{2}_{i=1} da_{i}ds_{i} \nonumber \\
& \hspace{30pt} \times \langle HH_{\beta}|\ell_{2}\rangle \langle\ell_{2}|\ell,a_{1},a_{2}\rangle \langle\ell,a_{1},a_{2}|\ell_{1}\rangle\langle \ell_{1}|HH_{\beta}\rangle, \label{4-1} \\ \nonumber
\end{align}
where $a_i$ and $s_i$ $(i=1,2)$ denote the sizes of the baby universes and the corresponding twist parameters, respectively. The $\ell$-geodesic and the $\ell_i$-geodesics $(i=1,2)$ are shown in Fig.\ref{fig7-1-g=2}.\\
\; \;  The principal challenge in evaluating (\ref{4-1}) is to determine the relevant integration regions in the moduli space of the genus-two geometry. The dependence on these moduli is encoded in the two transition amplitudes appearing in (\ref{4-1}). The amplitude $\langle \ell,a_1,a_2|\ell_1\rangle$ describes the simultaneous emission of two baby universes with sizes $a_1$ and $a_2$, whereas $\langle \ell_2|\ell,a_1,a_2\rangle$ describes their subsequent absorption. To evaluate the moduli-space integral, we therefore first formulate these transition amplitudes in terms of the Weil--Petersson volume. More generally, the transition amplitude for the emission of $n$ baby universes is given by\\
\begin{align}
\langle\ell,a_1,\dots,a_n|\ell_{1}\rangle = \int_{0}^{\infty}bdb \int_{0}^{\infty} dE \rho_{\text{Tr}}(E,b) V_{0,n+1}(b,a_1,\dots,a_n) \psi_{E}(\ell) \psi_{E}(\ell_{1}). \label{4-2} 
\end{align}
\vspace{-1.0em}
To evaluate this integral, we proceed as in the previous section. We first perform the inverse Laplace transform as in (\ref{3-3}) and parametrize the energies according to (\ref{3-4}):\\
\begin{align}
G(E,T)_{\chi=-3} &= e^{-3S_{0}} \int_{-\infty}^{\infty} e^{\ell}e^{-\Delta \ell} d\ell \int^{\infty}_{0} \prod_{i=1}^{2} b_idb_{i} \int \prod^{2}_{i=1} da_{i}ds_{i} \int^{\infty}_{-\infty} d\omega e^{-iT\omega} \ \rho_{\text{Tr}}(E_1,b_1) \rho_{\text{Tr}}(E_2,b_2) \nonumber \\
& \times \psi_{E_1}(\ell)\psi_{E_2}(\ell)V_{0,3}(b_1,a_1,a_2)V_{0,3}(b_2,a_1,a_2). \label{4-3} \\ \nonumber
\end{align}
\; \;  The dependence of (\ref{4-3}) on the relative sizes of the geodesic boundaries is encoded in the Weil--Petersson volumes $V_{0,3}(b_i,a_1,a_2)$. Since the moduli space for $(g,n)=(0,3)$ consists of a single point, the corresponding Weil--Petersson volume is $V_{0,3}(b_i,a_1,a_2)=1$. Nevertheless, to apply the strip approximation and the no-shortcut condition, it is useful to decompose this volume into contributions from Kontsevich graphs. This decomposition distinguishes the regions defined by the relative sizes of the boundary lengths $\{a_1,a_2,b_i\}$ and identifies the ribbon graph associated with each region. The integral in (\ref{4-3}) can then be reorganized as a finite sum of contributions from distinct strip geometries. Following the genus-one analysis of Ref.\cite{M.Sato}, we impose the no-shortcut condition on each configuration to determine the region in which the chosen geodesic defines the shortest admissible spatial slice. This procedure provides a tractable framework for evaluating a class of genus-two contributions to the two-point correlation function.
\vspace{-1.0em}
\subsection{Kontsevich graph for $V_{0,3}(b_i,a_1,a_2)$}
\begin{figure}[H]
\centering
\includegraphics[width=180mm]{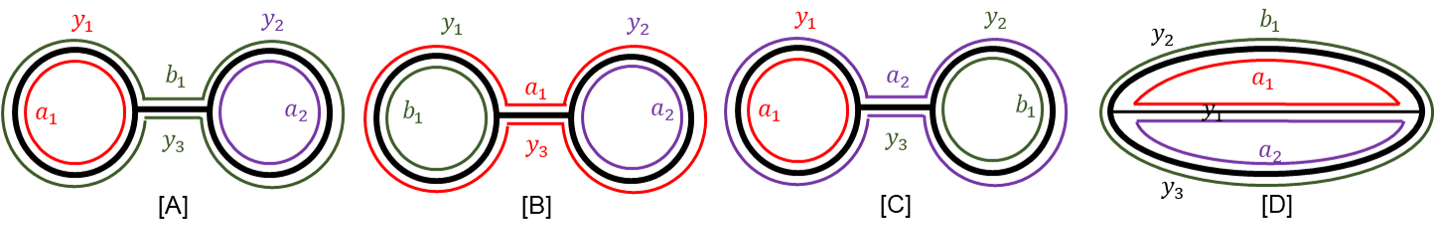}
\caption{\small The four trivalent Kontsevich graphs contributing to $V_{0,3} (b_1,a_1,a_2)$}
\label{fig7-6-2-Airy}
\end{figure}

\begin{figure}[H]
\centering
\includegraphics[width=140mm]{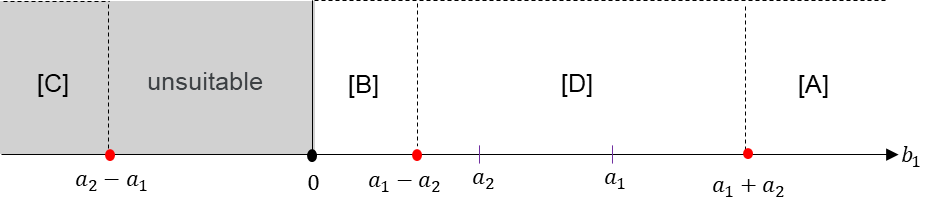}
\caption{\small  Regions of positive $b_1$ in which graphs [A], [B], and [D] contribute under the assumption $a_1>a_2$. Together, these regions cover the full range $0\leq b_1<\infty$.}
\label{fig7-7-strip0}
\end{figure}

\vspace{-0.5em}
\; \;  We begin with a brief review of the strip approximation. As discussed in the previous section, this approximation is motivated by the Gauss--Bonnet theorem. For a hyperbolic surface with geodesic boundaries, the area is fixed by the Euler characteristic. In the late-time regime, the relevant geodesic lengths become large while the area remains fixed. The surface is therefore well approximated by elongated strip-like regions and can be represented by a ribbon graph. This representation provides a tractable description of the relevant moduli-space integrals, even for higher-genus surfaces \cite{mulasepenner,norbury}.\\
\; \;  The ribbon-graph description is particularly useful in the limit of large boundary lengths $b_i$. In this limit, the leading asymptotic behavior of the Weil--Petersson volume $V_{g,n}(b_1,\dots,b_n)$ is described by the corresponding ribbon-graph volume, known as the Airy volume. The Airy volume is obtained by integrating over the edge lengths of the ribbon graph, subject to delta-function constraints that fix the length of each boundary to the prescribed value $b_i$. It can be expressed as a sum over Kontsevich graphs, namely the set $\Gamma_{g,n}$ of trivalent ribbon graphs of genus $g$ with $n$ boundaries, as follows \cite{Koncte-1,Koncte-2,ander}:\\
\begin{align}
V^{\text{Airy}}_{g,n}(\textbf{a})=\sum_{\Gamma\in\Gamma_{g,n}}\frac{2^{2g-2+n}}{\left|\text{Aut}\left(\Gamma \right)  \right| }\prod_{k=1}^{E}\int_{0}^{\infty}\text{d}y_{k}\prod_{i=1}^{n}\delta\left( a_i-\sum_{k=1}^{n}n^{i}_{k}y_{k}\right), \label{4-4}  \\ \nonumber
\end{align}
where $a_i$ is the length of the $i$-th geodesic boundary, $y_k$ is the length of the $k$-th edge of the graph, and $n_k^i \in \{0,1,2\}$ counts the number of times the $k$-th edge is incident on the $i$-th boundary. For $(g,n)=(0,3)$, the relevant Kontsevich graphs are classified into the four cases shown in Fig.\ref{fig7-6-2-Airy}.\\
\; \; Although the moduli space for $(g,n)=(0,3)$ consists of a single point, the contribution from each Kontsevich graph is supported in a different region determined by the relative sizes of the geodesic boundary lengths $b_1$, $a_1$, and $a_2$. The Weil--Petersson volume $V_{0,3}(b_1,a_1,a_2)$ can therefore be decomposed into graph contributions associated with these regions. We now evaluate each contribution using (\ref{4-4}). For graph [A], we obtain \\
\begin{align}
V_{0,3}^{\text{[A]}}(b_1,a_1,a_2)&=2\int_{0}^{\infty}\text{d}y_{1}\text{d}y_{2}\text{d}y_{3}\delta\left(b_1- y_1-y_2-2y_3\right)\delta\left(a_1- y_1\right) \delta\left( a_2-y_2\right)
\nonumber \\
&= \theta \left(b_1-a_1-a_2 \right)  \quad (b_1>a_1+a_2). \label{4-5} 
\end{align}
\\
The contributions from graphs [B] and [C] are obtained by cyclically permuting the geodesic boundary lengths $b_1$, $a_1$, and $a_2$:\\
\begin{align}
V_{0,3}^{\text{[B]}}(b_1,a_1,a_2)&= \theta \left(a_1-b_1-a_2\right) \quad(b_1<a_1-a_2), \label{4-6} \\
V_{0,3}^{\text{[C]}}(b_1,a_1,a_2)&= \theta \left(a_2-a_1-b_1\right)\quad(b_1<a_2-a_1). \label{4-7} 
\end{align}
\vspace{-1.0em}
\vspace{-0.5em}
\noindent \; \;  Finally, graph [D] corresponds to the region in which the three boundary lengths satisfy the triangle inequalities. Its contribution is symmetric under permutations of the boundary lengths and is given by\\
\begin{align}
V_{0,3}^{\text{[D]}}(b_1,a_1,a_2)&=2\int_{0}^{\infty}\text{d}y_{1}\text{d}y_{2}\text{d}y_{3}\delta\left( y_1+y_2-a_1\right)\delta\left(y_1+y_3- a_2\right) \delta\left( y_2+y_3-b_1\right)
\nonumber \\
&= \theta \left(a_1-a_2+b_1\right)\theta \left(a_2-a_1+b_1\right)\theta \left(a_1+a_2-b_1\right). \label{4-8} \\ \nonumber 
\end{align}
Summing the contributions from the four graphs, we obtain\\
\begin{align}
V_{0,3}(b_1, a_1, a_2) = \sum_{i=\text{[A]}}^{\text{[D]}} V_{0,3}^{i}(b_1, a_1, a_2) = 1. \label{4-9} 
\end{align}
\\
This result reproduces the known Weil--Petersson volume $V_{0,3}(b_1,a_1,a_2)=1$ and confirms the consistency of the Kontsevich-graph decomposition. We now apply this decomposition to the integral in (\ref{4-3}) in the regime $b_i,a_i,E\gg 1$. For $b_1>0$, graph [B] contributes when $a_1>a_2$, whereas graph [C] contributes when $a_2>a_1$. Since the expression is symmetric under the exchange $a_1\leftrightarrow a_2$, we may assume $a_1>a_2$ without loss of generality. Under this assumption, the entire range $0<b_1<\infty$ is divided among graphs [A], [B], and [D], as shown in Fig.\ref{fig7-7-strip0}.\\
\vspace{-1.0em}
\subsection{Evaluation Using the Strip Approximation}
\begin{figure}[H]
\centering
\includegraphics[width=185mm]{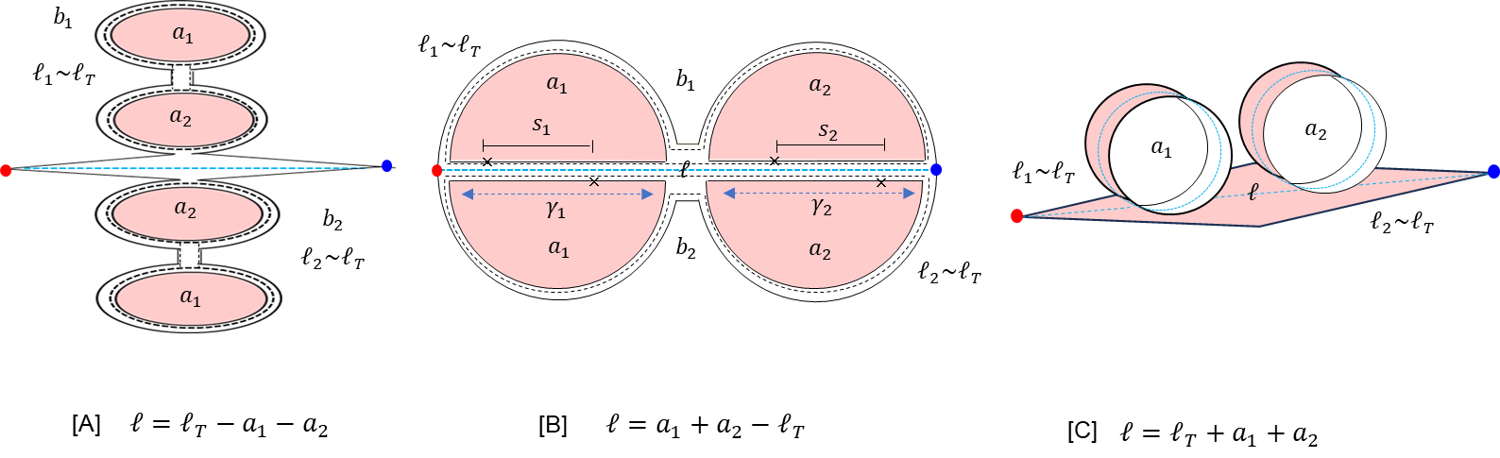}
\caption{\small Ribbon graphs corresponding to the geometric configurations associated with the delta functions in (\ref{4-14}). The two baby universes are represented by distinct closed geodesics. In configuration [A], the $\ell$-geodesic indicated by the dotted line defines the
shortest admissible spatial slice within the strip approximation. In configuration [B], the $\ell$-geodesic is the shortest candidate spatial
slice within the strip model described in Sec.~4.3, in the region defined by (\ref{4-21}); the explicit evaluation further assumes the symmetric
partition $\gamma_1=\gamma_2=\ell/2$. In configuration [C], the shorter of the $\ell_1$- and $\ell_2$-geodesics defines the shortest admissible spatial slice.}
\label{fig7-7-strip1}
\end{figure}

\; \; We now evaluate the contributions from graphs [A], [B], and [D], beginning with graph [A]. The support condition for this graph is $b_i>a_1+a_2$ for each $i=1,2$. The integral over each $b_i$ in this region is evaluated as\\
\begin{align}
\int_{a_1+a_2}^{\infty}db \ \frac{b \cos (b \sqrt{2E})}{\pi \sqrt{2E}} = -\frac{1}{\pi\sqrt{2E}}\Bigg[ \frac{A \sin (\sqrt{2E}A )}{\sqrt{2E}} + \frac{\cos (\sqrt{2E}A )}{2E} \Bigg] \approx -\frac{A \sin ( \sqrt{2E}A)}{2\pi E}+O(E^{-3/2}),  \label{4-10} 
\end{align}
\\
where $A$ is defined by $A=a_1+a_2$. In this evaluation, the following Fourier transform is used:\\
\begin{align}
\int^{\infty}_{0} db b^{k}\frac{\cos(b\sqrt{2E})}{\pi \sqrt{2E}} = -\frac{1}{\pi}(2E)^{-\frac{k}{2}-1} \Gamma(k+1)\sin\frac{\pi k}{2}. \label{4-11} \\ \nonumber
\end{align}
In the regime $a_i,E\gg 1$, the first term in (\ref{4-10}), which is proportional to the sine function, is of order $O(E^{-1})$. The second term, which is proportional to the cosine function, is subleading and of order $O(E^{-3/2})$. We therefore neglect the latter term in the large-$E$ approximation used throughout this section. Applying (\ref{4-10}) to both $b_i$ integrals, the contribution from graph [A] reduces to\\
\begin{align}
& G(E,T)_{\chi=-3}^{\text{[A]}}  \nonumber \\ 
& \approx \frac{e^{-3S_{0}}}{4\pi^2 E_1E_2} \int_{-\infty}^{\infty} e^{\ell}e^{-\Delta \ell} d\ell \int \prod^{2}_{i=1} da_{i}ds_{i} \int^{\infty}_{-\infty} d\omega e^{-iT\omega} A^{2} \sin(\sqrt{2E_1}A) \sin(\sqrt{2E_2}A) \psi_{E_1}(\ell)\psi_{E_2}(\ell). \label{4-12} 
\end{align}
\\
For sufficiently large $E_i$, applying the approximation in (\ref{3-6}) to $\psi_{E_i}(\ell)$ allows $G(E,T)_{\chi=-3}^{\text{[A]}}$ to be rewritten as \\
\begin{align}
G(E,T)_{\chi=-3}^{\text{[A]}} &\approx -\frac{e^{-3S_{0}-2\pi \sqrt{2E}}}{2\sqrt{2} E^{5/2}} \int_{-\log E}^{\infty} e^{-\Delta \ell} d\ell \int \prod^{2}_{i=1} da_{i}ds_{i} \int^{\infty}_{-\infty} \frac{d\omega}{2\pi} e^{-iT\omega} \nonumber \\
& \times A^2\left(e^{i A\left(\sqrt{2E}+\frac{\omega }{2\sqrt{2E}}\right)}-e^{-iA\left(\sqrt{2E}+\frac{\omega }{2\sqrt{2E}}\right)}\right) \left(e^{i A\left(\sqrt{2E}-\frac{\omega }{2\sqrt{2E}}\right)}-e^{-iA\left(\sqrt{2E}-\frac{\omega }{2\sqrt{2E}}\right)}\right)  \nonumber \\
& \times \left( -ie^{2i\sqrt{2E}L} + e^{i\frac{L}{\sqrt{2E}}\omega} + e^{-i\frac{L}{\sqrt{2E}}\omega} +ie^{-2i\sqrt{2E}L} \right). \label{4-13} \\ \nonumber 
\end{align}
The terms containing $e^{\pm 2i\sqrt{2E}L}$ or $e^{\pm 2iA\sqrt{2E}}$ oscillate rapidly and average to zero over a small energy window. After these terms are discarded, the $\omega$ integral yields delta functions:\\
\begin{align}
G(E,T)_{\chi=-3}^{\text{[A]}}  &\approx \frac{e^{-3S_{0}-2\pi \sqrt{2E}}}{2E^{2}} \int_{0}^{\infty} e^{-\Delta \ell} d\ell \int \prod^{2}_{i=1} da_{i}ds_{i} (a_1+a_2)^2 \nonumber \\
& \times \left[ \delta(\ell - \ell_T+ a_1 +a_2) + \delta(\ell + \ell_T-a_1 -a_2) + \delta(\ell -\ell_T -a_1 -a_2) \right]\theta(a_1-a_2). \label{4-14} 
\end{align}
\\
In deriving (\ref{4-14}), we neglect the subleading terms $\pm\log(2E)\mp 2$ appearing in the definitions of $L$ and $\ell_T$. Accordingly, we use $L\simeq\ell$ and $\ell_T\simeq\sqrt{2E}\,T$. We also replace the lower limit of the $\ell$ integration by zero, consistently with the large-$E$ approximation. The fourth delta function, $\delta(\ell+\ell_T+a_1+a_2)$, has no support in the region $\ell,\ell_T,a_1,a_2>0$ and has therefore been omitted from (\ref{4-14}). Applying the same procedure to graphs [B] and [D], we obtain\\
\begin{align}
G(E,T)_{\chi=-3}^{\text{[B]}}& \approx \frac{e^{-3S_{0}-2\pi \sqrt{2E}}}{2E^{2}} \int_{0}^{\infty} e^{-\Delta \ell} d\ell \int \prod^{2}_{i=1} da_{i}ds_{i} (a_1-a_2)^2 \nonumber \\
& \times \left[ \delta(\ell - \ell_T+ a_1 -a_2) + \delta(\ell + \ell_T-(a_1 -a_2)) + \delta(\ell -\ell_T -(a_1 -a_2)) \right]\theta(a_1-a_2), \label{4-15}
\end{align}
\begin{align}
&G(E,T)_{\chi=-3}^{\text{[D]}} \approx \frac{e^{-3S_{0}-2\pi \sqrt{2E}}}{2E^{2}} \int_{0}^{\infty} e^{-\Delta \ell} d\ell \int \prod^{2}_{i=1} da_{i}ds_{i} \nonumber \\
& \times \Big[ (a_1+a_2)^2 \left\{ \delta(\ell - \ell_T+ a_1 +a_2) + \delta(\ell + \ell_T-a_1 -a_2) + \delta(\ell -\ell_T -a_1 -a_2) \right\} \nonumber \\
& + (a_1-a_2)^2 \left\{ \delta(\ell -\ell_T+ a_1 -a_2) + \delta(\ell + \ell_T-(a_1 -a_2)) + \delta(\ell -\ell_T -(a_1 -a_2)) \right\} \Big]\theta(a_1-a_2). \label{4-16} 
\end{align}
\\
Unlike graph [A], graphs [B] and [D] contain delta functions involving the combination $a_1-a_2$. These terms describe configurations in which the two baby universes affect the ERB length in opposite ways.\\
\;\; Following the interpretation proposed in Ref.\cite{Firewall-1}, we regard the ERB length as a measure of the effective age of the parent universe for times below the Heisenberg time, $T_H=2\pi\rho(E)\sim O(e^{S_0})$. The emission of a baby universe of size $a>0$ rejuvenates the parent universe by shortening the ERB length according to $\ell_{\rm after}=\ell_{\rm before}-a$. By contrast, the absorption of a baby universe increases the ERB length and corresponds to the aging of the parent universe, $\ell_{\rm after}=\ell_{\rm before}+a$. For example, the condition imposed by $\delta(\ell-\ell_T+a_1-a_2)$ is $\ell=\ell_T-a_1+a_2$.
This relation describes a process in which a baby universe of size $a_1$ is emitted, while another baby universe of size $a_2$ is absorbed.\\
\;\; The delta functions in (\ref{4-16}) therefore encode the changes in the ERB length associated with the emission and absorption of two baby universes. Identifying the ERB lengths before and after these processes as $\ell_{\rm before}=\ell_T$ and $\ell_{\rm after}=\ell$, respectively, we classify the corresponding changes as follows:\\
\begin{align}
&\ell= \ell_T-(a_1+a_2) \ (\text{rejuvenation}), \hspace{10pt} \ell=-\ell_T+(a_1+a_2) \ (\text{aging}), \hspace{10pt} \ell=\ell_T+(a_1+a_2) \ (\text{aging}), \nonumber \\
&\ell= \ell_T-(a_1-a_2) \ (\text{rejuvenation}), \hspace{10pt} \ell=-\ell_T+(a_1-a_2) \ (\text{aging}), \hspace{10pt} \ell=\ell_T+(a_1-a_2) \ (\text{aging}). \label{4-17} 
\end{align}
\\
Identifying the relevant regions of moduli space for a genus-two Riemann surface is technically challenging. To address this difficulty, we apply the strip approximation following the genus-one analysis reviewed in the previous section. The ribbon graphs associated with the individual delta functions are then used to identify the relevant geometric configurations and their admissible integration regions \cite{M.Sato}.\\
\;\; In the following, we do not attempt to evaluate all the contributions listed in (\ref{4-17}). Instead, we focus on the three configurations involving the combination $a_1+a_2$. Among them, $\delta(\ell-\ell_T+a_1+a_2)$ describes the simultaneous emission of two baby universes, which shortens the ERB length of the parent universe by $a_1+a_2$. We analyze the corresponding ribbon graphs and determine which configurations satisfy the no-shortcut condition and contribute independently to the late-time two-point correlation function.
\vspace{-0.5em}
\paragraph{$\blacksquare$ \ $\delta(\ell-\ell_T+a_1+a_2)$ }
We now apply the strip approximation to the configuration associated with $\delta(\ell-\ell_T+a_1+a_2)$ in (\ref{4-14}). Cutting the two handles along the closed geodesics of lengths $a_1$ and $a_2$ yields the ribbon graph shown in Fig.\ref{fig7-7-strip1}[A]. In this graph, the closed geodesics of lengths $a_1$ and $a_2$ do not form the two boundaries of the same ribbon. Moreover, the ribbon traversed by the $\ell$-geodesic is not directly adjacent to either closed geodesic. Consequently, varying the twist parameters $s_1$ and $s_2$ does not generate a shortcut between two points on the $\ell$-geodesic. The $\ell$-geodesic therefore remains the shortest admissible spatial slice throughout the moduli-space region associated with this configuration, and the no-shortcut condition imposes no additional restriction.\\
\;\; The twist parameters are integrated over $0<s_1<a_1$ and $0<s_2<a_2$, a representative range for the identifications $s_i\sim s_i+a_i$.  For fixed $\ell$, the delta function imposes the constraint $a_1+a_2=\ell_T-\ell$. Together with $a_1>0$, $a_2>0$, and $a_1>a_2$, this constraint implies $\frac{1}{2}(\ell_T-\ell)<a_1<\ell_T-\ell$, with $a_2=\ell_T-\ell-a_1$. Equivalently, $a_2$ lies in the range $0<a_2<\frac{1}{2}(\ell_T-\ell)$. Using these integration domains, the corresponding contribution to the two-point correlation function is given by\\
\begin{align}
&G(E,T)_{\chi=-3}^{\text{[A]},\ell=\ell_T -a_1-a_2}    \nonumber \\
&= \frac{e^{-3S_{0}-2\pi \sqrt{2E}}}{2E^{2}}  \int_{0}^{\ell_T} e^{-\Delta \ell}d\ell   \int^{\ell_T-\ell}_{\frac{1}{2}(\ell_T-\ell)} da_1 \int^{\frac{1}{2}(\ell_T-\ell)}_{0} da_2 \int^{a_1}_{0} ds_1 \int^{a_2}_{0} ds_2 (a_1+a_2)^2 \delta(\ell-\ell_T+a_1+a_2) \nonumber \\
& =  \frac{e^{-3S_{0}-2\pi \sqrt{2E}}}{24E^{2}} \int^{\ell_T}_{0} e^{-\Delta \ell}(\ell_T-\ell)^{5} d\ell \nonumber \\
& = \frac{e^{-3S_0-2\pi\sqrt{2E}}}{24\Delta E^2} \Bigg[\ell_T^5-\frac{5\ell_T^4}{\Delta}+\frac{20\ell_T^3}{\Delta^2}-\frac{60\ell_T^2}{\Delta^3} + \frac{120\ell_T}{\Delta^4}-\frac{120}{\Delta^5}+\frac{120e^{-\Delta \ell_T}}{\Delta^5} \Bigg]. \label{4-18}
\end{align}
\vspace{-0.5em}
\noindent \;  Consider the limit $\Delta \to 0$ in (\ref{4-18}). The factor $e^{-\Delta\ell}$ represents the matter contribution associated with the two operator insertions and depends on the length $\ell$ of the chosen spatial slice. As $\Delta \to 0$, this factor approaches unity, and the operator-dependent weight disappears. The remaining integral then reduces to the gravitational path integral over the region of moduli space selected by the no-shortcut condition. In this sense, the $\Delta \to 0$ limit of (\ref{4-18}) can be interpreted as a partial contribution to the partition function.\\
\;\; This procedure should be distinguished from the first method presented in Ref.\cite{M.Sato}, in which one sums over all homotopically inequivalent geodesics connecting the two boundary operators. In that approach, the weight associated with each geodesic approaches $e^{-\Delta\ell_\gamma}\to 1$ as $\Delta\to 0$. The sum over geodesics therefore becomes divergent. Consequently, the partition function cannot be obtained directly from (\ref{2-8}) by simply taking the $\Delta\to 0$ limit.\\
\;\; Taking the limit $\Delta\to 0$ in (\ref{4-18}), we obtain the following partial contribution to the partition function $Z_{2,1}(\beta)$:\\
\begin{align}
\lim_{\Delta \to 0} G(E,T)_{\chi=-3}^{\text{[A]},\ell=\ell_T -a_1-a_2} &= \frac{e^{-3S_{0}-2\pi \sqrt{2E}}}{18}ET^6 \nonumber \\ \underset{\text{Laplace transform}}{\rightarrow} Z_{2,1}^{\text{[A]},\ell=\ell_T -a_1-a_2}(\beta) & = \frac{e^{-3S_{0}}T^6}{9\beta^2} \left[ \frac{\pi^2}{\beta} - \frac{3\sqrt{2\pi^3}}{4\beta^{1/2}}e^{\frac{2\pi^2}{\beta}}\left( 1+\frac{4\pi^2}{3\beta} \right) \operatorname{Erfc} \left( \sqrt{\frac{2\pi^2}{\beta}}\right) + \frac{1}{2} \right]. \label{4-19}
\end{align} 
\paragraph{$\blacksquare$ \ $\delta(\ell+\ell_T-a_1-a_2)$ } The geometric configuration associated with $\delta(\ell+\ell_T-a_1-a_2)$ is represented by the ribbon graph shown in Fig.\ref{fig7-7-strip1}[B]. The analysis below is restricted to this strip configuration.
\paragraph{Setup} In this graph, the closed geodesics $a_1$ and $a_2$ lie on two separate ribbons, which are joined only at the narrow central region of the ribbon graph between the labels $b_1$ and $b_2$ in Fig.\ref{fig7-7-strip1}[B]; we call this region the neck. The $\ell$-geodesic passes through the neck and crosses the $i$-th ribbon along a subinterval of length $\gamma_i$, with $\gamma_1+\gamma_2=\ell$. The twist parameter $s_i$ identifies two points on the $i$-th ribbon separated by $s_i$. We treat $\gamma_i$ as a quantity characterizing the ribbon graph rather than as an additional modulus. After imposing the delta-function constraint, the remaining independent moduli are $a_1$, $s_1$, and $s_2$. For the explicit evaluation below, we impose the symmetric partition $\gamma_1=\gamma_2=\ell/2$. This relation is not derived from the edge lengths of the graph in Fig.\ref{fig7-7-strip1}[B] and, since $a_1\neq a_2$ in general, does not follow from a symmetry of the configuration. It should therefore be regarded as an additional simplifying prescription. For an asymmetric partition, the allowed region in the $(s_1,s_2)$ plane, and consequently the integrand in (\ref{4-22}), would change. The present result is therefore restricted to the symmetric prescription $\gamma_1=\gamma_2=\ell/2$. A determination of $\gamma_i$ is left for future work.
\vspace{-1.0em}
\paragraph{Local path comparison in the strip model}In the strip representation of Fig.\ref{fig7-7-strip1}[B], the twist parameter $s_i$ identifies the two sides of the $i$-th ribbon with a relative displacement $s_i$ along the cutting curve $a_i$. Within the metric ribbon-graph model, this identification gives two local alternatives to the reference segment of length $\gamma_i$, with lengths $s_i$ and $a_i-s_i$, respectively. This is the same local path comparison as in the genus-one strip configuration of Fig.\ref{fig5-2(strip)}[B], with $(b,\tau,\ell)$ replaced by $(a_i,s_i,\gamma_i)$.\\
\;\; Throughout, we call a path \emph{retained} if it is one of the candidate paths included in the strip model. These consist of the two outer paths $\ell_1$ and $\ell_2$, together with the neck-crossing paths obtained by choosing, independently in each ribbon, one of the three local segments of lengths $\gamma_i$, $s_i$, and $a_i-s_i$. The reference arc is the neck-crossing path corresponding to the choice  $\gamma_i$ in both ribbons. Paths outside this class are not considered.\\
\;\; Closing either of the two local alternatives through the identification defines an auxiliary closed curve on the genus-two surface. For any such auxiliary curve $c_i$, the following parity constraint applies. For each pair of pants in (4.3), the total number of transverse intersections of a closed curve with its three boundary components must be even. Consequently, if $c_i$ intersects $a_i$ once and is disjoint from $a_j$, $j\neq i$, then the parity condition requires additional intersections with the remaining cutting curves. In particular, $c_i$ cannot be disjoint from both $b_1$ and $b_2$.\\
\;\; We therefore do not use a Dehn twist about any such auxiliary curve. We neither assume that $b_1$ and $b_2$ are preserved by such a twist nor use the twist to define a transformation of the genus-two cutting data. Instead, the lengths $a_i$, $s_i$, and $\gamma_i$ are compared directly within the fixed metric ribbon graph of Fig.\ref{fig7-7-strip1}[B].\\
\;\; The standard Dehn-twist identification about the cutting curve $a_i$ acts on the Fenchel--Nielsen twist parameter as 
\begin{equation*}
  s_i\longmapsto s_i+a_i . 
\end{equation*}
It leaves the pants-curve length coordinates unchanged. We use this periodic identification only to choose the representative
\begin{equation*}
  0<s_i<a_i ,
\end{equation*}
as in (\ref{4-18}). No claim is made here about how the lengths of curves or arcs intersecting $a_i$ transform under this twist; the lengths of the retained candidate arcs are read off directly from the fixed metric ribbon graph.\\
\;\; The reference segment in the $i$-th ribbon is shorter than both local alternatives if and only if \\
\begin{equation*}
  \gamma_i<s_i \qquad\text{and}\qquad \gamma_i<a_i-s_i,  
\end{equation*}
or equivalently, 
\begin{align}
  \gamma_i<s_i<a_i-\gamma_i . \label{4-20}
\end{align}
\\
Because the two ribbons are connected only through the neck, a retained neck-crossing path traverses the two ribbons successively. The restriction to the two local alternatives in each ribbon, as well as the assumption that the local choices in the two ribbons may be made independently, is part of the strip model adopted here. Neither property is derived from the genus-two mapping class group. Under these assumptions, the length of a retained neck-crossing path is the sum of its two local contributions.
\vspace{-1.0em}
\paragraph{Minimality within the strip model}We now examine whether the reference arc of length $\ell$ is minimal within the strip model of Fig.\ref{fig7-7-strip1}[B]. In this model, the surface is represented by the corresponding metric ribbon graph. A candidate spatial slice is represented by a path on this graph connecting the two marked points, and its length is approximated by the total length of the corresponding edge path.\\
By the definition above, the retained candidate arcs are the two outer paths $\ell_1$ and $\ell_2$, whose lengths are approximated by
\begin{equation*}
  \ell_1\simeq\ell_2\simeq\ell_T=a_1+a_2-\ell ,
\end{equation*}
and the neck-crossing paths, which traverse ribbon~1 and ribbon~2 successively. Within ribbon $i$, the reference segment of length
$\gamma_i$ is compared with the two local alternatives of lengths $s_i$ and $a_i-s_i$. Paths coupling the two handles in a manner not represented by this metric ribbon graph are not included in the present analysis.\\
\;\; Under the assumption that the local choices in the two ribbons are independent, the minimum length among the retained neck-crossing paths is the sum of the individual minima in the two ribbons. Since
\begin{equation*}
  \gamma_1+\gamma_2=\ell ,
\end{equation*}
the reference arc is the shortest among the retained neck-crossing paths if and only if the reference segment is the shortest in each ribbon. (\ref{4-20}) for $i=1,2$ gives\\
\begin{align}
  \gamma_i<s_i<a_i-\gamma_i , \qquad i=1,2 .\label{4-21}
\end{align}
\\
The interval in (\ref{4-21}) is nonempty only when $a_i>2\gamma_i$. Under the symmetric prescription $\gamma_i=\ell/2$, this condition becomes $a_i>\ell$. \\
\;\; At leading order in the strip approximation, the outer paths have lengths $\ell_1,\ell_2\simeq\ell_T$. In the integration range $0<\ell<\ell_T$ adopted in (\ref{4-22}), the outer paths are therefore longer, at this order, than the reference arc of length $\ell$. Thus, under the assumptions stated above, (\ref{4-21}) specifies the region in which the reference arc is the shortest among the candidate spatial slices retained in this strip model.\\
\;\; Away from the boundaries of this region, where two or more retained candidate arcs have equal length, the shortest representative within the retained class is unique. These boundaries form a set of measure zero and therefore do not affect the integral.\\
\;\; For the explicit evaluation, we further impose the symmetric prescription $\gamma_1=\gamma_2=\ell/2$. We do not claim that this prescription is preserved by any element of the mapping class group, that the region (\ref{4-21}) is a fundamental domain for the full mapping class group action, or that it counts every geometry represented by Fig.\ref{fig7-7-strip1}[B] exactly once. Accordingly, (\ref{4-22}) gives the contribution from this restricted subfamily within the strip model, rather than the complete contribution of all geometries in the branch associated with $\delta(\ell+\ell_T-a_1-a_2)$. More precisely, it should be understood as a partial, prescription-dependent
contribution to the genus-two sector of the two-point function (\ref{2-4}), evaluated within the strip model (with its restricted class
of candidate slices) under the symmetric prescription $\gamma_1=\gamma_2=\ell/2$. Within the present strip-model prescription, its identification with the full genus-two contribution to (\ref{2-4}) would at least require a treatment of the more general partition $\gamma_1+\gamma_2=\ell$, together with a proof that the relevant integration domain accounts for each geometry exactly once.
It would further require establishing, at genus two, the equivalence between the slice-selection prescription used here and the sum over geodesics (\ref{2-8}), as well as the evaluation of the remaining branches in (\ref{4-17}).\\
\;\; The argument does not establish minimality on the full genus-two moduli space. It also does not apply to the other branches in (\ref{4-17}) or to the configurations shown in Fig.\ref{fig7-7-strip1}[A] and Fig.\ref{fig7-7-strip1}[C].
\paragraph{Integration domain} The conditions in (\ref{4-21}) are imposed separately for each $i$ and are not equivalent to the single condition obtained by summing them. Summing (\ref{4-21}) over $i$ gives $\ell<s_1+s_2<a_1+a_2-\ell$. This is only a necessary condition and defines an overly large domain, namely a strip in the $(s_1,s_2)$ plane. We therefore do not use it. With the symmetric partition $\gamma_1=\gamma_2=\ell/2$, the individual conditions define the rectangle $ \frac{\ell}{2}<s_i<a_i-\frac{\ell}{2},\quad( i=1,2)$. This rectangle is non-empty only for $a_1,a_2>\ell$. The twist integrals therefore give its area,\\
\[
\int_{\ell/2}^{a_1-\ell/2}ds_1 \int_{\ell/2}^{a_2-\ell/2}ds_2 =(a_1-\ell)(a_2-\ell). 
\]\\
The remaining integration domain is determined by \\
\[ a_1+a_2=\ell+\ell_T, \qquad a_1>a_2, \qquad a_1,a_2>\ell.
\]\\
Using the delta function to eliminate $a_1$ and setting $x=a_2-\ell$, we find \\
\[ a_2=\ell+x, \qquad a_1-\ell=\ell_T-\ell-x.
\]\\
The condition $a_1>a_2$ then gives $0<x<\frac{\ell_T-\ell}{2}$. Hence, \\
\begin{align}
&\int_0^\infty da_1 \int_0^\infty da_2\, (a_1-\ell)(a_2-\ell)\, \delta(\ell+\ell_T-a_1-a_2) \theta(a_1-a_2) \theta(a_1-\ell) \theta(a_2-\ell) \nonumber\\ &= \int_0^{(\ell_T-\ell)/2} x(\ell_T-\ell-x)\,dx = \frac{(\ell_T-\ell)^3}{12}. \nonumber 
\end{align}
\\
On the support of the delta function, $(a_1+a_2)^2=(\ell_T+\ell)^2$. The contribution is therefore\\ 
\begin{align}
G(E,T)&_{\chi=-3}^{\text{[A]},\ell= a_1+a_2-\ell_T} = \frac{e^{-3S_{0}-2\pi \sqrt{2E}}}{2E^{2}}  \int_{0}^{\ell_T} e^{-\Delta \ell} d\ell \int^{\infty}_{0} da_1 \int^{\infty}_0 da_2   \int_{\ell/2}^{a_1-\ell/2} ds_1 \int_{\ell/2}^{a_2-\ell/2} ds_2 (a_1+a_2)^2 \nonumber \\
&\times  \delta(\ell+\ell_T-a_1-a_2)\theta(a_1-a_2)\theta(a_1-\ell)\,\theta(a_2-\ell) \nonumber \\
&=  \frac{e^{-3S_{0}-2\pi \sqrt{2E}}}{24E^{2}} \int_{0}^{\ell_T} e^{-\Delta \ell}(\ell_T+\ell)^{2}(\ell_T-\ell)^3 d\ell \nonumber \\
&= \frac{e^{-3S_0-2\pi\sqrt{2E}}}{24\Delta E^2} \Bigg[ \left( \frac{24\ell_T^2}{\Delta^3} + \frac{96\ell_T}{\Delta^4}+\frac{120}{\Delta^5} \right)e^{-\Delta \ell_T}+ \ell_T^5 - \frac{\ell_T^4}{\Delta} - \frac{4\ell_T^3}{\Delta^2} +\frac{12\ell_T^2}{\Delta^3} +\frac{24\ell_T}{\Delta^4}-\frac{120}{\Delta^5} \Bigg]. \label{4-22}
\end{align}
The region $a_2<\ell$ does not contribute because the individual conditions in (\ref{4-21}) make the corresponding rectangle empty. By contrast, the strip domain obtained from the necessary condition alone would receive a nonvanishing contribution from this region.\\
\;\; Taking the limit $\Delta\to0$ in (\ref{4-22}), we obtain the corresponding partial contribution to the partition function $Z_{2,1}(\beta)$:\\
\begin{align}
\lim_{\Delta \to 0} G(E,T)_{\chi=-3}^{\text{[A]},\,\ell=a_1+a_2-\ell_T} &= \frac{11e^{-3S_{0}-2\pi \sqrt{2E}}}{90}ET^6 \nonumber \\ \underset{\text{Laplace transform}}{\rightarrow} Z_{2,1}^{\text{[A]},\ell= a_1+a_2-\ell_T}(\beta) &= \frac{11e^{-3S_{0}}T^6}{45\beta^2} \left[ \frac{\pi^2}{\beta} - \frac{3\sqrt{2\pi^3}}{4\beta^{1/2}}e^{\frac{2\pi^2}{\beta}}\left( 1+\frac{4\pi^2}{3\beta} \right) \operatorname{Erfc} \left( \sqrt{\frac{2\pi^2}{\beta}}\right) + \frac{1}{2} \right]. \label{4-23} 
\end{align}
\paragraph{$\blacksquare$ \ $\delta(\ell-\ell_T-a_1-a_2)$ }Finally, we consider the ribbon graph shown in Fig.\ref{fig7-7-strip1}[C], which represents the configuration associated with $\delta(\ell-\ell_T-a_1-a_2)$. Within the strip model for this configuration, there is no region in which the $\ell$-geodesic satisfies the no-shortcut condition. Within the class of candidate slices considered here, the shortest admissible slice connecting the two boundary points is therefore the shorter of the two outer paths $\ell_1$ and $\ell_2$, rather than the $\ell$-geodesic. Consequently, this configuration does not provide an independent leading genus-two contribution within the present prescription. Instead, it gives a correction to the disk contribution, in the same way as the contribution in (\ref{3-14}). \\
\\
In this section, we examined the strip configurations associated with the combination $a_1+a_2$. For the configuration shown in Fig.\ref{fig7-7-strip1}[B], we classified the candidate spatial slices included in the strip model into the outer paths and the arcs passing through the neck. Under the assumptions of this model, we showed that the $\ell$-geodesic is the shortest among these candidates in the region defined by (\ref{4-21}). For the explicit calculation, we further imposed the symmetric partition $\gamma_1=\gamma_2=\ell/2$. The resulting integration domain therefore describes only a restricted subfamily within the strip model. We do not identify this domain with a fundamental domain of the full genus-two mapping class group. Accordingly, the result represents the contribution from this restricted
subfamily. It is a partial, prescription-dependent contribution to the genus-two sector of (\ref{2-4}), and should not be interpreted as the complete contribution of the branch associated with $\delta(\ell+\ell_T-a_1-a_2)$ or as the complete genus-two contribution.

\section{Conclusion}
\; \; Our analysis suggests that the Kontsevich-graph decomposition, the strip approximation, and the no-shortcut condition together provide a useful analytic framework for studying selected higher-genus configurations. Within the strip approximation, this framework allows us to identify and evaluate contributions from specified regions of the strip parameter space without directly integrating over the full moduli space. The present analysis is restricted to the genus-two branches involving the combination $a_1+a_2$. Even within these branches, the explicit result is obtained only from a specified region of the strip parameter space. It is not the complete integral over the genus-two moduli space modulo the full mapping class group. Accordingly, the result (4.22) is a partial, prescription-dependent contribution to the genus-two sector of the two-point function (\ref{2-4}), and its relation to the full genus-two term remains to be established.\\
\; \; Establishing global minimality would require a classification of all admissible slices generated by the full genus-two mapping class group, including transformations that couple the two handles, together with a proof that the resulting integration domain intersects each relevant mapping-class-group orbit exactly once. These questions, as well as a first-principles determination of $\gamma_i$, lie beyond the present analysis. A complete genus-two analysis would also require the evaluation of the remaining emission and absorption processes, including those involving $a_1-a_2$. A quantitative comparison with the higher-order spectral correlations predicted by random matrix theory would provide an important test of the present framework. Extending the framework to arbitrary genus may further clarify the role of topology change in the late-time regime of quantum gravity and its relation to unitarity.\\
\; \; A physical interpretation suggested by the strip configuration of (\ref{4-18}), which describes the simultaneous emission of two baby universes, is that topology-changing processes can partially reverse the growth of the black hole interior. In this configuration the ERB length is reduced by $a_1+a_2$. This produces a contribution that is not exponentially suppressed by the semiclassical growth of the ERB. Crucially, however, this late-time effect is not determined by topology alone. Within the strip approximation, the no-shortcut condition selects the region of the strip parameter space in which the chosen $\ell$-geodesic is the shortest among the candidate spatial slices retained in the model. Late-time correlations can therefore receive contributions from geometric configurations that counteract the semiclassical growth of the ERB. This result suggests that the relevant late-time physics is controlled not only by genus, but also by the geodesic structure of the contributing configurations.

\end{document}